\documentclass[%
 reprint,
superscriptaddress,
 amsmath,amssymb,
 aps,
prb,
]{revtex4-2}

\usepackage{graphicx}% Include figure files
\usepackage{dcolumn}% Align table columns on decimal point
\usepackage{bm}% bold math
\usepackage{amsmath}
\usepackage{braket}
\usepackage{hyperref}% add hypertext capabilities
\usepackage[utf8]{inputenc}
\usepackage{newunicodechar}
\newunicodechar{∼}{\textasciitilde}
\usepackage[mathlines]{lineno}% Enable numbering of text and display math

\begin{document}

%\preprint{APS/123-QED}

\title{Layerwise Stratification and Band Reordering in Twisted Multilayer 
MoTe\textbf{$\rm{_2}$}}
\thanks{*These two authors contribute equally to this work.}%

\author{Yueyao Fan*}
\affiliation{Department of Materials Science and Engineering, University of Washington, Seattle, WA 98195, USA}
\author{Xiao-Wei Zhang*}
\affiliation{Department of Materials Science and Engineering, University of Washington, Seattle, WA 98195, USA}
\author{Yusen Ye}
\affiliation{Department of Materials Science and Engineering, University of Washington, Seattle, WA 98195, USA}
\author{Xiaoyu Liu}
\affiliation{Department of Materials Science and Engineering, University of Washington, Seattle, WA 98195, USA}
\author{Chong Wang}
\affiliation{Department of Materials Science and Engineering, University of Washington, Seattle, WA 98195, USA}
\author{Kaijie Yang}
\affiliation{Department of Materials Science and Engineering, University of Washington, Seattle, WA 98195, USA}
\author{Di Xiao}
\email{dixiao@uw.edu}
\affiliation{Department of Materials Science and Engineering, University of Washington, Seattle, WA 98195, USA}
\affiliation{Department of Physics, University of Washington, Seattle, WA 98195, USA}
\author{Ting Cao}
\email{tingcao@uw.edu}
\affiliation{Department of Materials Science and Engineering, University of Washington, Seattle, WA 98195, USA}

\begin{abstract}
We introduce a generalizable, physics‑informed strategy for generating training data that enables a machine‑learning force field accurate over a broad range of twist angles and stacking layer numbers in moir\'e systems.
Applying this to multilayer twisted MoTe\textbf{$\rm{_2}$} (tMoTe$_2$), we identify a structural and electronic stratification: the two moir\'e interface (MI) layers retain substantial lattice reconstruction even in thick multilayers, while outer bulk-like layers show rapidly attenuated distortions.
Surprisingly, this stratification becomes strongest not in the ultra-small twist angle regime ($\lesssim 1$\textdegree), where in-plane domain formation is well known, but rather at intermediate angles ($2–5$\textdegree).
Simultaneously, interlayer hybridization across the MI‑bulk boundary is strongly suppressed, leading to electronic isolation.
In twisted double bilayer MoTe\textbf{$\rm{_2}$}, this stratification gives rise to coexisting honeycomb and triangular lattice motifs in the frontier valence bands.
We further demonstrate that twist angle and weak gating can create energy shift of bands belonging to the two motifs, producing Chern band reordering and nonlinear electric polarization with modest hole doping.
Our approach allows efficient simulation of multilayer moir\'e systems and reveals structural‑electronic separation phenomena absent in bilayer systems.
\end{abstract}

\maketitle

Moir\'e superlattices provide a versatile platform for discovering new structural and electronic behaviors in solid state\cite{Nuckolls2024-wk}. 
Their emergent properties, from fractional Chern insulators\cite{Cai2023-oo, Zeng2023-tq, Park2023-hz, Xu2023-vq, Lu2024-nf, Xie2021-bs, Aronson2025-yt, Park2025-an, Waters2025-lx, Dong2025-gz, Morales-Duran2024-bi} to moir\'e ferroelectricity\cite{zheng2020unconventional, yasuda2021stacking, weston2022interfacial, ding2024engineering}, are intimately tied to changes in the underlying lattice structure induced by stacking, reconstruction, or symmetry-breaking distortions.
Moir\'e stacking universally induces structural relaxation, which reshapes atomic-scale displacements across a wide range of systems, from graphene \cite{yoo2019atomic} to transition-metal dichalcogenide (TMD) homo- and heterostructures\cite{weston2020atomic, rosenberger2020twist, van2023rotational}.
In TMD moir\'e systems, this lattice response is further enriched by strong coupling between mechanical deformations and internal polarization fields, giving rise to emergent topological textures and twist-angle–driven Chern-number reversal, as supported by both theoretical\cite{Zhang2024-zh} and experimental studies\cite{Cai2023-oo, foutty2024mapping}.
Beyond twist angle, the number of stacking layers introduces another powerful degree of freedom to engineer moir\'e structures: in thick multilayers, layers distant from the moir\'e interface (MI) can acquire stacking motifs reminiscent of bulk polytypes, such as 2H and 3R in TMD or rhombohedral and Bernal stacking in graphene, creating mixed-dimensional architectures that combine moir\'e-interface and quasi-bulk behaviors, as demonstrated in twisted $1+n$ graphene systems\cite{waters2023mixed} and multilayer graphene on BN\cite{Lu2024-nf, mullan2023mixing, lu2025extended, xie2025tunable}.

Despite this progress, how reconstruction and electronic coupling evolve across layers in multilayer moir\'e superlattices remains incompletely understood, especially in the intermediate twist-angle regime (2–$5^\circ$) most relevant for correlated and topological phases in systems like twisted MoTe\textbf{$\rm{_2}$} (tMoTe$_2$)\cite{Cai2023-oo, Park2023-hz}. 
Continuum elasticity suggests that increasing layer thickness would suppress reconstruction since thicker layers are mechanically stiffer, and that any strain or displacement would smoothly decay away from the twisted interface\cite{halbertal2023multilayered, cazeaux2020energy, halbertal2021moire}.
But such descriptions do not capture the distinct stacking energetics between moir\'e layers, moir\'e–bulk interfaces, and bulk-like regions, nor the coupled in-plane and out-of-plane strain propagation shaped by both intralayer elasticity and the stacking energetics. 
A quantitative, atomistic description, serving as a foundation for understanding electronic properties, requires simulations that are both large-scale and stacking-registry aware. 
Machine-learning force fields (MLFFs) provide the needed scalability and accuracy, yet prior MLFF studies have largely focused on bilayers, where small bilayer supercells or unit cells suffice to provide training data set\cite{Zhang2024-zh, Liu2025-jw, Shaidu2025-su, Georgaras2025-hk}. 
Multilayer systems such as $n+m$ tMoTe$_2$ introduce additional R- and H-type interfaces and out-of-plane relaxation that extend beyond adjacent layers, enlarging the configurational space well beyond bilayers and motivating an approach that generalizes across angles and layer numbers.

\begin{figure*}
\center
    \includegraphics[width=\textwidth]{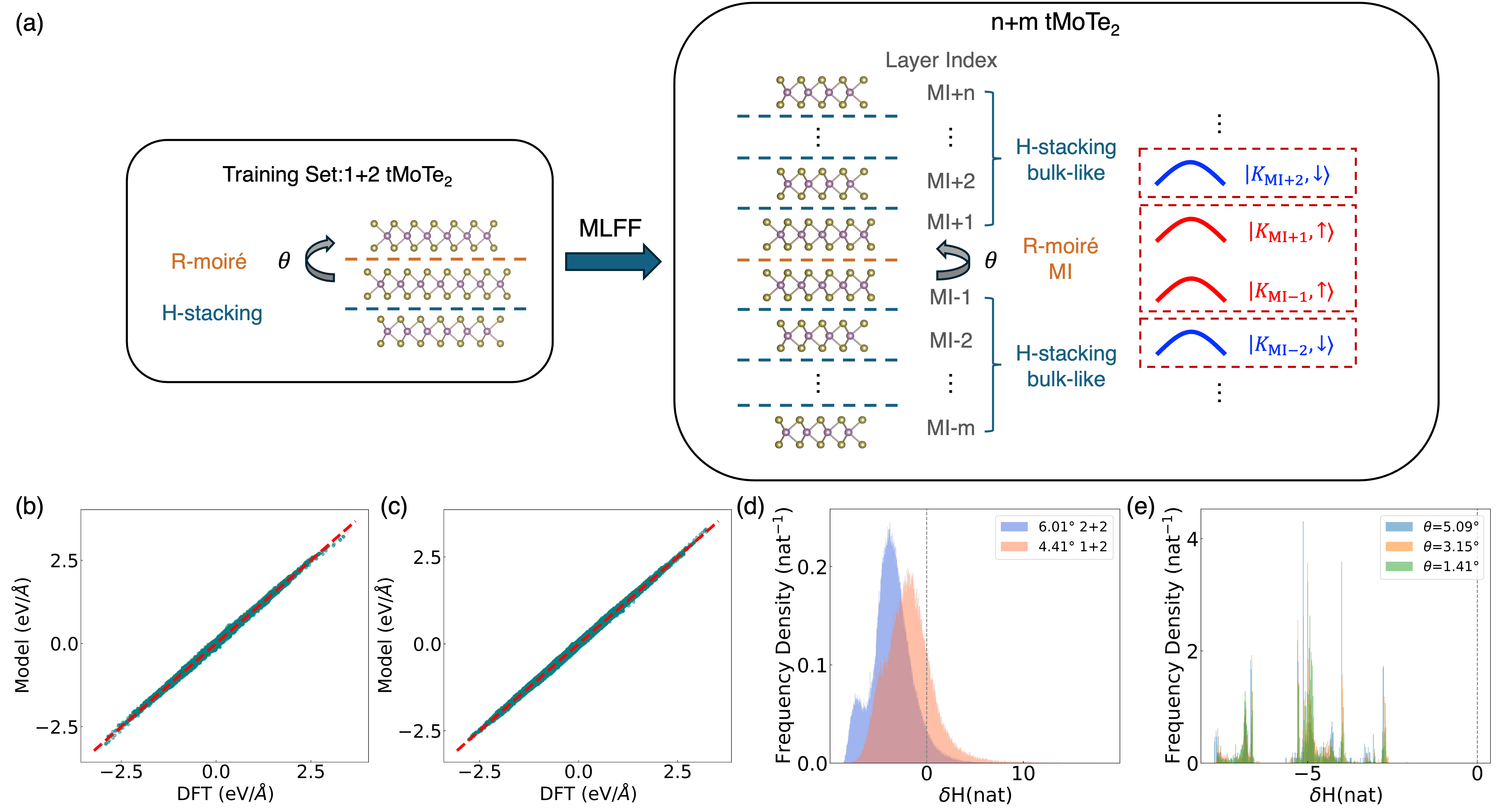}
    \caption{\label{Figure1} 
    (a) Schematic structure of $n+m$ tMoTe$_2$ with twist angle $\theta$ near R stacking. Layers are indexed as MI$-m$ to MI$+n$. Inset: (Right) layer-resolved bands color-coded by layer polarization with red at MI and blue at bulk-like layers. $\ket{K_{\text{MI}+j}, \uparrow/\downarrow}$ shows the monolayer state at Brillouin zone corner K of the layer $\text{MI}+j$ with spin $\uparrow/\downarrow$. (Left) $1+2$ tMoTe$_2$ for training data generation. (b) Force parity plot for MLFF tested on $4.41^\circ$ $1+2$ tMoTe$_2$, comparing the MLFF-model predicted and DFT calculated ionic forces on each atom (green dots). Red line of $y=x$ shows conditions of exact match. (c) Force parity plot for $6.01^\circ$ $2+2$ tMoTe$_2$. (d, e) Differential information entropy distribution of local atomic environment for (d) $4.41^\circ$ $1+2$ (red), $6.01^\circ$ $2+2$ (blue), and (e) $5+5$ tMoTe$_2$ under $\theta=5.09$\textdegree\ (blue), $3.15$\textdegree\ (orange), and $1.41$\textdegree\ (green) relative to $1+2$ training set. Frequency density is the normalized distribution of $\delta H$ datapoints over the full range.}
\end{figure*}

Here we develop an efficient training data generation strategy for studying $n+m$ tMoTe$_2$ (Fig.~\ref{Figure1}(a)) by capturing the complete layer interfaces configurations with $1+2$ tMoTe$_2$ and obtain a MLFF generalizable for twist angles and stacking layers.
In multilayer tMoTe$_2$, we observe structural and electronic layer stratification that are unexpected from continuum elasticity theory.
The system stratifies into two motifs, MI layers and bulk-like outer layers.
The atomic displacement at MI dominates over that in bulk-like layers at intermediate twist angles, and is robust against increasing layer thickness, indicating layer structural segregation.
Electronically, the frontier valence bands are isolated due to negligible interlayer tunneling across MI-bulk boundary.
Demonstrated in $2+2$ tMoTe$_2$, the orbitals in MI form a hexagonal pattern and those in outer layers form triangular ones, respectively.
The bands from the two motifs show different valley Chern numbers and can flip order in energy by varying twist angles.
Furthermore, at modest hole doping, a weak out-of-plane electric field can induce a nonlinear polarization by shifting the holes between the two motifs.

\textit{Layerwise transferable MLFF-}
The lattice reconstruction of $n+m$ multilayer tMoTe$_2$ across twist angles could be predicted by MLFF trained on $1+2$ tMoTe$_2$ at a single twist angle which contains a near-complete set of layer interface configurations. 
The key principle is that MLFF uses local atomic descriptors that encode the embedded local environment of each atom, including both its intralayer coordination and the registry with adjacent layers.
As long as all relevant local atomic environments are present in the training set, the model can be reliably transferred to larger and thicker moir\'e systems. 

Fig.~\ref{Figure1}(a) shows the structure of $n+m$ multilayer tMoTe$_2$, where $n$ layer-2H stacking MoTe$_2$ is twisted on the top of $m$ layers of 2H stacking MoTe$_2$ with an angle $\theta$, forming a R-type MI in between.
Counting from the layer above MI, the layers are labeled by $\text{MI}+1$, $\text{MI}+2$, ..., $\text{MI}+n$ and the layers below are denoted as $\text{MI}-1$ to $\text{MI}-m$.  

Assuming the lattice reconstruction in moir\'e systems is mainly decided by interactions between adjacent layers \cite{Wang2018-wp, Zeng2023-lv, Batzner2022-ud}, the distinct types of lattice interface configuration in $n+m$ tMoTe$_2$ are the one R-type MI and multiple 2H interfaces in bulk-like layers. 
Thus, we could generate the training set of MLFF on one R-type MI and one 2H interface, with the minimal system containing both being $1+2$ tMoTe$_2$ (Fig.~\ref{Figure1}(a)).
Compared to training separately on a bilayer moir\'e R-type and a bilayer 2H system, the $1+2$ structure provides advantages: (1) it is balanced between the two types of interfaces and easy to prepare, and (2) it partially incorporates the effect of interactions beyond nearest-neighbor layers, making it more representative for multilayer systems.
We then generate training data on $1+2$ tMoTe$_2$ at $\theta=6.01$\textdegree\ by running 5500 steps of ab initio molecular dynamics (AIMD) at 500K and randomly sample $80\%$ of the snapshots for training (see Supplementary Information).

Our trained MLFF can accurately generate lattice reconstruction patterns in $n+m$ tMoTe$_2$ across various twist angles and layer thickness. 
To prove MLFF's transferability across twist angles, we test its force predictions on a set of trajectory of $4.41$\textdegree\ $1+2$ tMoTe$_2$ against DFT-calculated forces. 
Fig.~\ref{Figure1}(b) shows the corresponding parity plot, where the predicted forces align closely with DFT values.
The model’s strong performance is further quantified by a force root-mean-square error (RMSE) below $0.03$ eV/\AA.
We also validate the MLFF’s transferability across layer thicknesses using $6.01$\textdegree\ $2+2$ tMoTe$_2$. 
Parity plot in Fig.\ref{Figure1}(c) shows agreement between MLFF predictions and DFT results, yielding a force RMSE below $0.04$ eV/\AA.

The MLFF's transferability roots in the completeness of the training set, which could be quantified by the negative differential information entropy, $\delta H$, a measure of how closely the local atomic environments in the target system match those in the training data\cite{LORDI2025}. Fig.\ref{Figure1}(d) shows the distribution of $\delta H$ for local atomic environment in the AIMD calculations of  $4.41$\textdegree\ $1+2$ tMoTe$_2$ and $6.01$\textdegree\ $2+2$ tMoTe$_2$, with $>80\%$ and $>95\%$ of datapoints exhibiting negative values, respectively (see Supplementary Information).

This high coverage of atomic environment in the training set enables the MLFF to be reliably extended to large, multilayer structures beyond the computational reach of direct DFT relaxation.
Fig.\ref{Figure1}(e) shows the $\delta H$ distribution of relaxed $5+5$ tMoTe$_2$ structures at $\theta=$ 1.41\textdegree , 3.15\textdegree\ and 5.09\textdegree\ with $>99\%$ of local environment having negative $\delta H$, confirming the training set completeness and indicating reliable MLFF predictions for thick-layer systems.

\begin{figure}
\center
\hspace{-10pt} 
    \includegraphics[width=\linewidth]{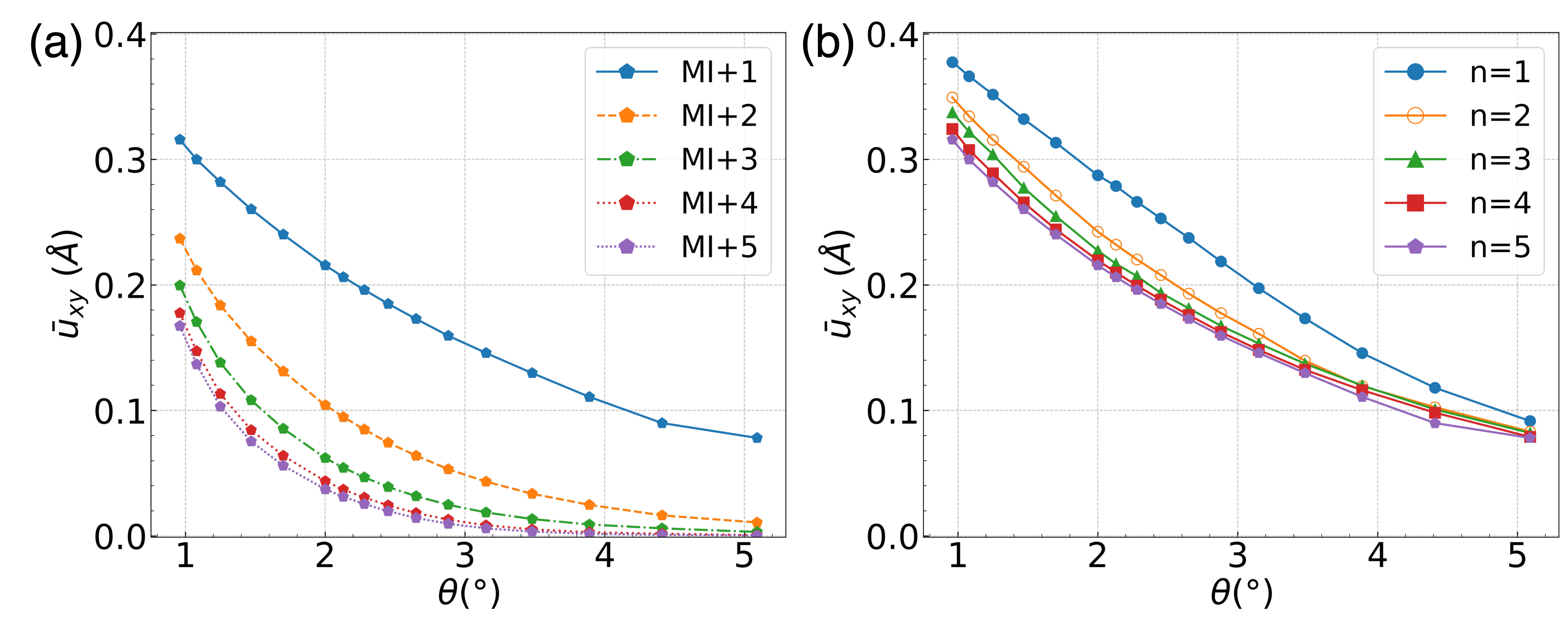}
    \caption{\label{Figure2} (a) The average magnitude of the Mo in-plane displacement $\bar{u}_{xy}$ across twist angles and layer indices in $5+5$ tMoTe$_2$. (b) $\bar{u}_{xy}$ at MI layers across twist angles and thickness in $n+n$ ($n=1, 2, ..., 5$) tMoTe$_2$. Lines are guides to the eye.}
\end{figure}

\textit{Structural stratification-}
The in-plane atomic displacement shows stratification between MI layers and bulk-like layers.
Fig.\ref{Figure2}(a) shows the average magnitude of layerwise Mo in-plane atomic displacement ($\bar{u}_{xy}$) across $\theta$ obtained by MLFF for the layers above MI in $5+5$ tMoTe$_2$, while the layers below MI have the same $\bar{u}_{xy}$ due to an in-plane two-fold rotational symmetry.
The in-plane atomic displacement patterns reveal that reconstruction displacement vectors belonging to each layer exhibit characteristic winding patterns around three-fold symmetric axes that switch helicity when crossing MI (see Supplementary Information).
Despite the similarity in the displacement pattern, the MI layers, i.e., $\text{MI}\pm1$, hold the largest $\bar{u}_{xy}$ with a nearly constant drop of 0.1~\AA\ to the bulk-like layers across the twist angle range of 2-5\textdegree.
The drop is much larger than the attenuation of $\bar{u}_{xy}$ across the bulk-like layers from $\text{MI}+2$ to $\text{MI}+5$, indicating a segregation between the R-type MI and the H-type bulk-like layers.
This sharp reconstruction gradient from MI$\pm 1$ to MI$\pm 2$ layers contradicts the na\"ive expectation of gradual strain propagation according to the elasticity of a continuum medium. 

Surprisingly, the stratification becomes strongest at intermediate angles ($2–5$\textdegree), rather than at smaller twist angle of $\lesssim1$\textdegree, where moir\'e reconstruction becomes significant and in-plane domain formation is observed \cite{weston2020atomic, quan2021phonon,weston2022interfacial,van2023rotational, tilak2022moire}. 
For example, at $3.89$\textdegree\ $5+5$ tMoTe$_2$, $\bar{u}_{xy}$ at the MI+1 layer is $4.5$ times of that at the MI+2 layer. 
In comparison, at $0.96$\textdegree, this ratio drops to $1.3$. 
These trends are consistent with a picture in which the shorter moir\'e period and absence of extended domains at $2$–$5$\textdegree\ tend to limit stress transfer from MI$\pm1$ to MI$\pm2$, whereas at $\lesssim 1$\textdegree\ the enlarged moir\'e cells and domain networks provide longer strain propagation paths, so MI$\pm2$ displacements can approach those at MI$\pm1$.

The in-plane atomic displacement at MI layers is robust against increasing layer thickness for all twist angles.
As shown in Fig.~\ref{Figure2}(b), when one 2H stacking layer is added at each side to $1+1$ tMoTe$_2$, the $\bar{u}_{xy}$ at MI layers in $2+2$ tMoTe$_2$  keeps $>85\%$ of the $\bar{u}_{xy}$ at $1+1$ tMoTe$_2$.
When layer thickness increases from $n=2$ to $5$ in $n+n$ tMoTe$_2$, the $\bar{u}_{xy}$ curves are almost overlapping, indicating that $\bar{u}_{xy}$ plateaus.
As such, the small effect from bulk-like layers to MI$\pm1$ layers are dominated by $\text{MI}\pm2$, while that of the disjoint layers is even smaller.
This further validates our training data generation strategy that 1+2 tMoTe$_2$ composing of both moir\'e R-type and bilayer 2H-type interfaces provides complete description of atomic environment in thick moir\'e layers. 

\begin{figure}
\includegraphics[width=\columnwidth]{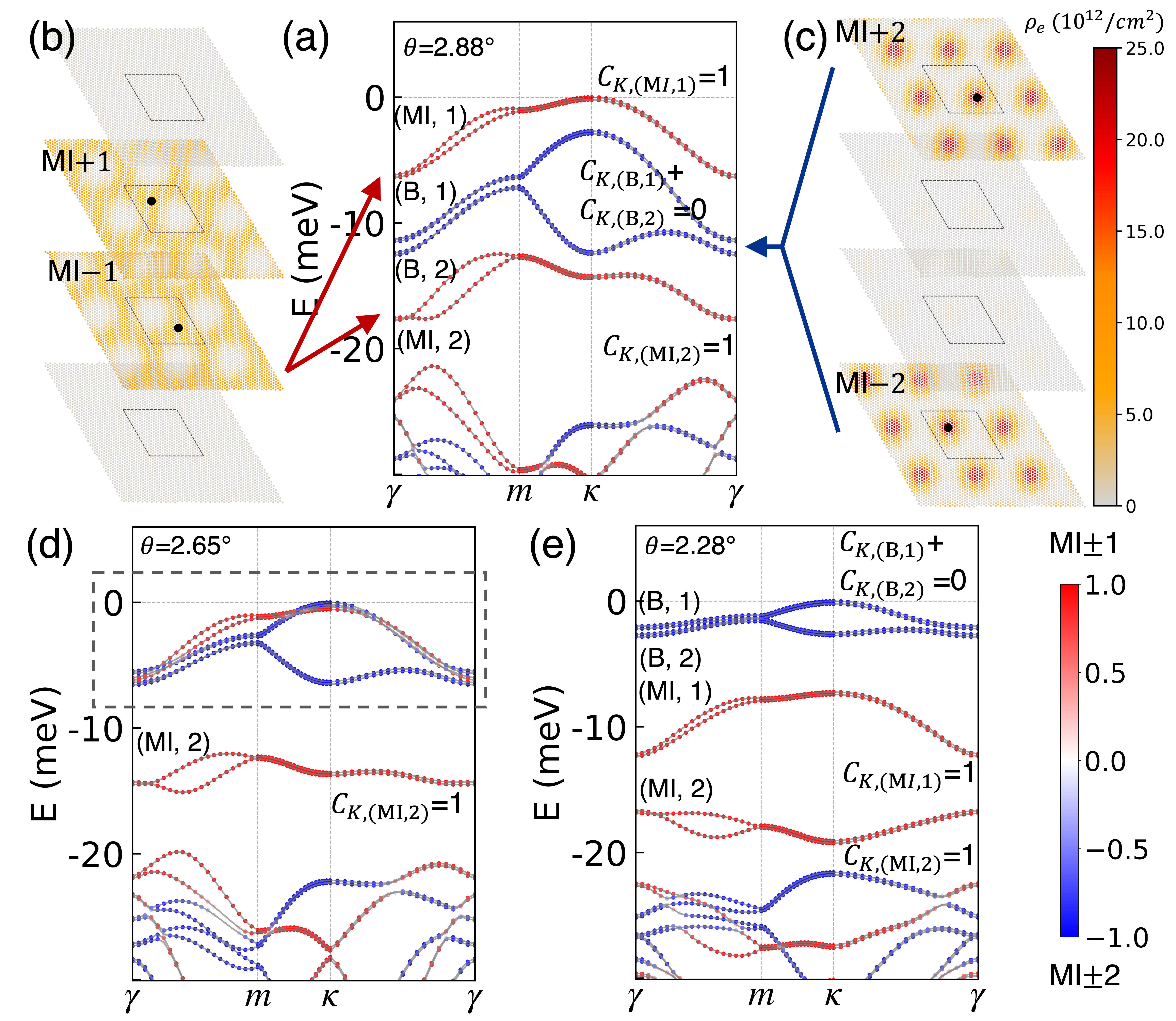}% Here is how to import EPS art
\caption{\label{Figure3} 
(a) Band structure of $2.88$\textdegree\ $2+2$ tMoTe$_{2}$ with color coded layer polarization $S(\alpha, \mathbf{k})$ (defined in main text). $C_{K, \alpha}$ labels the valley Chern number.
(b, c) Electron density $\rho_e$ of states at $\gamma$, integrated along z and projected onto 2D. (b) $|\psi_{(\text{MI}, 1),\gamma}|^2+|\psi_{(\text{MI}, 2),\gamma}|^2$ and (c)$|\psi_{(\text{B}, 1),\gamma}|^2+|\psi_{(\text{B}, 2),\gamma}|^2$ for $2.88$\textdegree\ $2+2$ tMoTe$_{2}$. The density maxima in each layer are marked by black dots. The moir\'e cell is marked by black dashed boxes. 
(d, e) Valence bands of (d) $2.65$\textdegree\ and (e) $2.28$\textdegree\ $2+2$ tMoTe$_2$. Dashed box in (d) marks overlapping bands $(\text{MI}, 1)$, $(\text{B}, 1)$, and $(\text{B}, 2)$.
}
\end{figure}

\textit{Electronic stratification and Chern band reordering-}
In the following, we focus on the electronic structures of a minimal system that exhibits structural stratification, $2+2$ tMoTe$_2$.
The frontier orbitals in $2+2$ tMoTe$_2$ is located in either the MI layers ($\text{MI}\pm1$) or outer layers ($\text{MI}\pm2$) with little interlayer tunneling across the 2H interface in between.
Fig.\ref{Figure3}(a) shows the valence bands of $2.88$\textdegree\ $2+2$ tMoTe$_2$ with layer polarization $S(\alpha, \mathbf{k}) = \sum_{i} |\phi_{i,\alpha,\mathbf{k}}|^2l_i$. 
$\phi_{i,\alpha,\mathbf{k}}$ is the projection coefficient of the eigenstate $\psi_{\alpha,\mathbf{k}}$ for band $\alpha$ and momentum $\mathbf{k}$ onto atomic orbital $i$ (each $\alpha$ represents a set of two-valley degenerate bands, see Supplementary Information). 
The layer polarization is taken as $l_i= -1$ if $i$ belongs to atoms in $\text{MI}\pm2$, or $l_i= 1$, if belonging to $\text{MI}\pm1$.
Our calculations reveal clear electronic stratification, with $S(\alpha,\mathbf{k}) > 0.9$ or $< -0.9$ for the first four frontier bands, all strongly layer-polarized due to weak interlayer hybridization across the MI–bulk interface (red dashed box in Fig.~\ref{Figure1}(a)).

This layerwise electronic stratification manifests in coexisting motifs.
Fig.\ref{Figure3}(b) shows the $\gamma$ point electron density of bands $(\text{MI}, 1)$ and $(\text{MI}, 2)$ projected to each layer.
Positions of peaks density reveal that electrons distributed at MI $\pm 1$ layers combine together to form a hexagonal electronic lattice pattern.
For bands $(\text{B},1)$ and $(\text{B},2)$, as shown in Fig.\ref{Figure3}(c), electrons mostly localize at the outer layers that are spatially separated by MI $\pm 1$ layers.
The peak of the electron density at each layer forms a triangular lattice pattern.
As such, the negligible interlayer tunneling among $\text{MI}-2$, $\text{MI}\pm1$, and $\text{MI}+2$ leads to coexistence of two lattice motifs in the frontier valence bands: one honeycomb in $(\text{MI}, 1)$ -- $(\text{MI, 2})$, and two triangular in $(\text{B}, 1)$ and $(\text{B}, 2)$.

Decreasing twist angle introduces Chern bands reordering of the states belonging to the two lattice motifs in $2+2$ tMoTe$_2$.
When decreasing $\theta$ from $2.88$\textdegree\ (Fig.~\ref{Figure3}(a)) to $2.65$\textdegree\ (Fig.~\ref{Figure3}(d)), the energy of band $(\text{B}, 1)$ and $(\text{B}, 2)$ lifts and almost overlaps with $(\text{MI}, 1)$ over the moir\'e Brillouin zone.
Further reducing $\theta$ to $2.28$\textdegree\ (Fig.~\ref{Figure3}(e)) brings $(\text{B}, 1)$ and $(\text{B}, 2)$ to the top valence bands with a global gap over $(\text{MI}, 1)$, leading to reordering of bands from the two lattice motifs.

Meanwhile, bands $(\text{MI}, 1)$ and $(\text{MI}, 2)$ both carry valley Chern number $C_{K}=1$ in a wide range of $2-3$\textdegree, while bands $(\text{B}, 1)$ and $(\text{B}, 2)$ overlap in energy and show combined trivial topology.
Thus, the reordering with decreasing twist angle also leads to a change of valley Chern number of the top valence bands.
Notably, $(\text{MI}, 2)$ forms a flat Chern band with a bandwidth of $3$ meV (Fig.~\ref{Figure3}(d)) at $2.65^\circ$, larger than the $2.0^\circ$, where the second valence band in $1+1$ tMoTe$_2$ attains optimal flatness\cite{Zhang2024-zh, Wang2025-ov}. 
This upward shift improves experimental accessibility of the correlated topological phases at higher bands by mitigating twist-angle disorder and enhancing interaction effects via reduced moir\'e site spacing \cite{Wang2025-ov, Park2025-an, kang2024double}.

\begin{figure}
\includegraphics[width=\columnwidth]{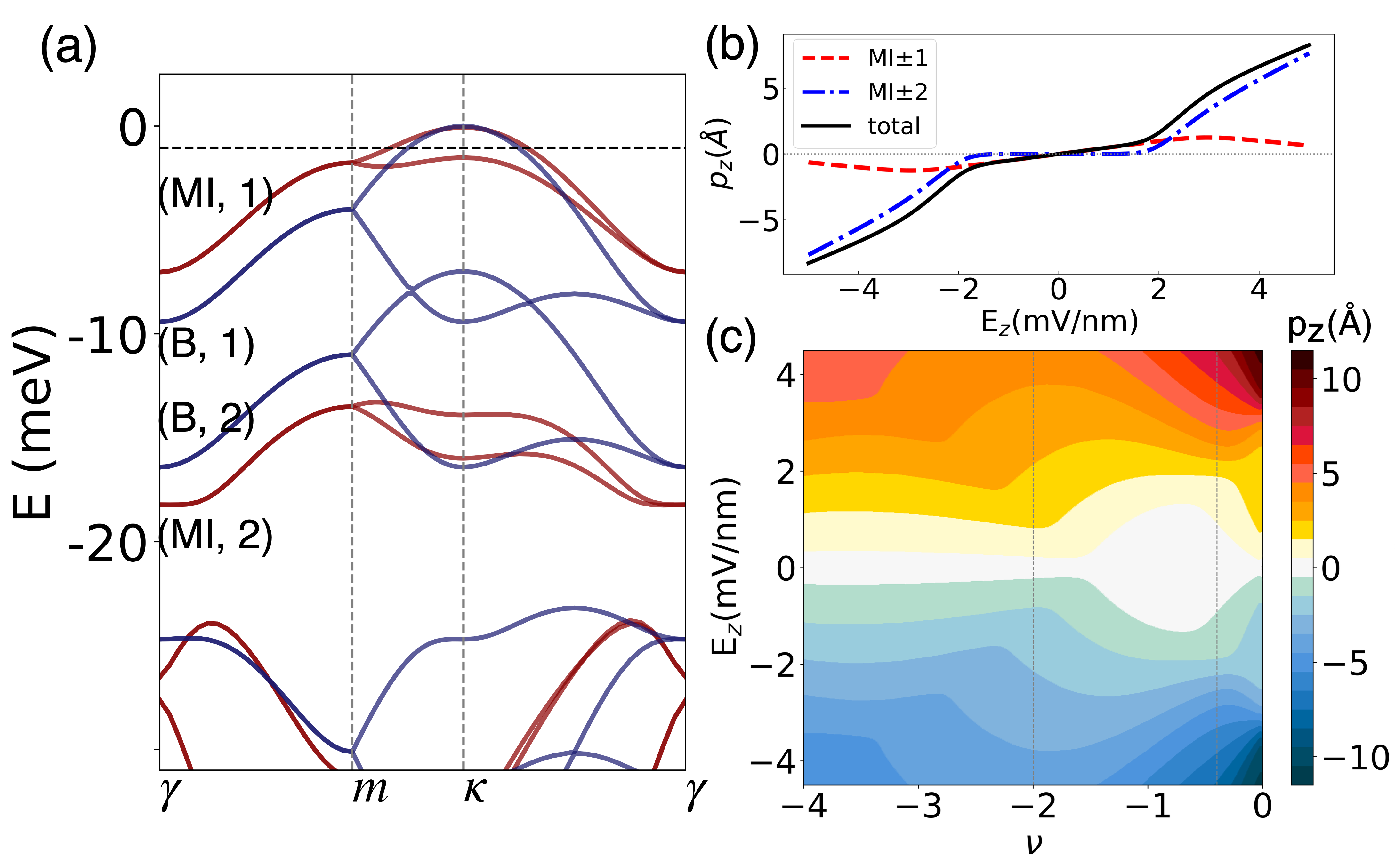}
\caption{\label{Figure4} 
(a) Valence bands of $2.88$\textdegree\ $2+2$ tMoTe$_2$ under $E_z=-3.3$ mV/nm. Red and blue bands are layer polarized at $\text{MI}\pm1$ and $\text{MI}\pm2$, respectively. A chemical potential corresponding to filling factor $\nu=-0.4$ is marked by black dashed line.
(b) The response of out-of-plane electric polarization per hole $p_z$ to out-of-plane electric field $E_z$ at hole filling $\nu=-0.4$. The total value, contributions from $\text{MI}\pm1$ and $\text{MI}\pm2$ are marked with black solid, red dashed, and blue dashed lines, respectively.
(c) Color map of the total electric polarization per hole $p_z$ as a function of hole filling factors and out-of-plane electric fields.}
\end{figure}

\textit{Gate-tunable nonlinear electric polarization-}
The layer polarization at valence band maxima (VBM) is also tunable by out-of-plane electric field $E_z$ causing band reordering.
We calculate the band structure of $2+2$ tMoTe$_2$ for a range of twist angles and under $E_z$, using continuum models fitted to DFT bands (see Supplementary Information). 
At $2.88$\textdegree, a very weak field of $E_z=-3.3$ mV/nm (corresponding to an interlayer potential difference of 2.3 meV) drives the bands $(\text{B}, 1)$ and $(\text{MI}, 1)$ into near degenerate at $\kappa$ (Fig.~\ref{Figure4}(a)).
With slightly stronger fields, band $(\text{B}, 1)$ shifts above $(\text{MI}, 1)$, switching the VBM layer polarization from the MI layers (Fig.~\ref{Figure3}(a)) to outer layers (Fig.~\ref{Figure4}(a)).

The electrically driven band reordering introduces nonlinear electric polarization upon light hole doping.
At a doping level $\nu=-0.4$, Fig.~\ref{Figure4}(b) shows the out-of-plane electric polarization per hole, $p_z$, which exhibits a turning point at $|E_z|\sim2.0$ mV/nm.
Here, $p_z=\sum_{j, \alpha, \mathbf{k}}|\tilde\psi_{j, \alpha, \mathbf{k}}|^2f(\epsilon_{\alpha, \mathbf{k}})\bar{z_j}/(\nu N)$, where $j$ indexes layer $\text{MI}+j$, $\tilde\psi_{j, \alpha, \mathbf{k}}$ is the projection of eigenstates on layer $\text{MI}+j$, $f(\epsilon_{\alpha, \mathbf{k}})$ is the Fermi-Dirac distribution, $\epsilon_{\alpha, \mathbf{k}}$ is the eigen energy, $\bar{z_j}$ is the average z coordinate of layer $\text{MI}+j$, and $\nu N$ is the number of doped holes. 
At small $|E_z|$, $p_z$ is dominated by contributions from $\text{MI}\pm1$, consistent with hole occupation of band $(\text{MI},1)$.
Beyond the turning point, $p_z$ follows contributions from $\text{MI}\pm2$, reflecting hole transfer into $(\text{B},1)$.
The abrupt redistribution of charge between electronically isolated layers produces the nonlinear $p_z$ response.
The turning point $|E_z|$ can be further reduced at a slightly smaller twist angle. 

Unlike in $1+1$ moir\'e systems where the layer polarization arises from continuous modulation of interlayer wavefunction weights due to field-induced interlayer potential bias, the nonlinear response in $2+2$ tMoTe$_2$ is governed by discrete hole transfer between electronically isolated layers, driven by band reordering under electric field.
This switch, accompanied by a transition from the honeycomb-like lattice motif to the triangular ones, occurs under an order-of-magnitude smaller fields than in $1+1$ tMoTe$_2$ \cite{anderson2023programming}, highlighting a strong layer–band selectivity and the sensitive coupling between topology and out-of-plane electric polarization.

The nonlinearity of electric polarization persists in a boarder range of doping level.
Fig.~\ref{Figure4}(c) is a map of $p_z$ over varied doping levels and out-of-plane electric fields, with Fig.~\ref{Figure4}(b) as the vertical dashed line at $\nu=-0.4$.
Besides, at the dashed line $\nu=-2$, we also observed nonlinearity originated from the band reordering between $(\text{B}, 1)$ and $(\text{B}, 2)$ (see Supplementary Information for details).

\textit{Conclusions-}
We introduce a minimal-layer training protocol that enables a single MLFF to generalize across layer counts and twist angles.
This approach uncovers robust layerwise structural and electronic stratification absent in bilayer moir\'e, leading to Chern band reordering and gate-tunable nonlinear polarization.
The coexistence of distinct lattice motifs suggests a broader landscape of emergent phases, where Chern bands in the honeycomb MI layers may be further reshaped by charge or magnetic ordering, as well as screening environments in the surrounding triangular lattices upon doping \cite{Li2021, Xu2020, Regan2020, Tang2020, Wang2020, Ghiotto2021, Zhou2021, Li2021_2}.  
Our finding provides new physical mechanisms to engineer emergent moir\'e physics in twisted multilayers.

\begin{acknowledgments}
This work is supported by the U.S. Department of Energy, Office of Basic Energy Sciences, under Contract No. DE-SC0025327. The development of machine-learning enabled methods and advanced codes was supported by the Computational Materials Sciences Program funded by the U.S. Department of Energy, Office of Science, Basic Energy Sciences, Materials Sciences, and Engineering Division, PNNL FWP 83557. 
Y.F was supported by the U.S.-Japan University Partnership for Workforce Advancement and Research and Development in Semiconductors (UPWARDS) . 
This research used resources of the National Energy Research Scientific Computing Center, a DOE Office of Science User Facility supported by the Office of Science of the U.S. Department of Energy under Contract No. DE-AC02-05CH11231 using NERSC award BES-ERCAP0032546 and BES-ERCAP0033256.
This work was also facilitated through the use of advanced computational, storage, and networking infrastructure provided by the Hyak supercomputer system and funded by the University of Washington Molecular Engineering Materials Center at the University of Washington (DMR-2308979). 
\end{acknowledgments}

\bibliography{apssamp}% Produces the bibliography via BibTeX.

\end{document}